\documentclass[%
 reprint,
superscriptaddress,
 amsmath,amssymb,
 aps,
]{revtex4-2}

\usepackage{graphicx}
\usepackage{dcolumn}
\usepackage{bm}
\usepackage{booktabs}
\begin{document}

\preprint{APS/123-QED}

\title[Interface-Driven Peptide Folding: Quantum Computations on Simulated Membrane Surfaces]{Interface-Driven Peptide Folding: Quantum Computations on Simulated Membrane Surfaces}

\author{Daniel Conde-Torres}
\affiliation{Departamento  de  Física  Aplicada, Facultade  de  Física, Universidade  de  Santiago  de Compostela,Santiago de Compostela, E-15782,Spain}%
\affiliation{Organic  Chemistry  Department, Centro  Singular  de  Investigación  en  Química  Biolóxica  e Materiais Moleculares (CiQUS), Universidade de Santiago de Compostela, Campus Vida s/n, Santiago de Compostela, E-15782, Spain}

\author{Mariamo Mussa-Juane}%
\affiliation{Centro de Supercomputación de Galicia (CESGA),Santiago de Compostela,15705, Spain}

\author{Daniel Faílde}%
\affiliation{Centro de Supercomputación de Galicia (CESGA),Santiago de Compostela,15705, Spain}

\author{Andrés Gómez}%
\affiliation{Centro de Supercomputación de Galicia (CESGA),Santiago de Compostela,15705, Spain}

\author{Rebeca García-Fandiño}
\email[E-mail me at: ]{rebeca.garcia.fandino@usc.es}
\affiliation{Organic  Chemistry  Department, Centro  Singular  de  Investigación  en  Química  Biolóxica  e Materiais Moleculares (CiQUS), Universidade de Santiago de Compostela, Campus Vida s/n, Santiago de Compostela, E-15782, Spain}
\affiliation{MD.USE Innovations S.L., Ed. Emprendia, Campus Vida s/n, Santiago de Compostela, E-15782, Spain}

\author{Ángel Piñeiro}
\email[E-mail me at: ]{angel.pineiro@usc.es}
\affiliation{Departamento  de  Física  Aplicada, Facultade  de  Física, Universidade  de  Santiago  de Compostela,Santiago de Compostela, E-15782,Spain}%
\affiliation{MD.USE Innovations S.L., Ed. Emprendia, Campus Vida s/n, Santiago de Compostela, E-15782, Spain}%

\date{\today}

\begin{abstract}
Antimicrobial peptides (AMPs) are pivotal in combating infections. They play important roles in broader health contexts, including cancer, autoimmune diseases, and aging. A critical aspect of AMP functionality is their targeted interaction with pathogen membranes, which often possess altered lipid compositions. Designing AMPs with enhanced therapeutic properties relies on a nuanced understanding of these interactions, which are believed to trigger a rearrangement of these peptides from random coil to alpha-helical conformations, essential for their lytic action. Traditional supercomputing has consistently encountered difficulties in accurately modeling these structural changes, especially within membrane environments, thereby opening an opportunity for more advanced approaches. This study extends an existing quantum computing algorithm, initially designed to simulate peptide folding in homogeneous environments, by adapting it to address the complexities of antimicrobial peptide (AMP) interactions at interfaces. Our approach enables the prediction of the optimal conformation of peptides located in the transition region between hydrophilic and hydrophobic phases, akin to lipid membranes. The new method has been applied to model the structure of three 10-amino-acid-long peptides, each exhibiting hydrophobic, hydrophilic, or amphipathic properties in different media and at interfaces between solvents of different polarity. Notably, our approach does not demand a higher number of qubits compared to simulations in homogeneous media, making it more feasible with current quantum computing resources. Despite existing limitations in computational power and qubit accessibility, our findings demonstrate the significant potential of quantum computing in accurately characterizing complex biomolecular processes, particularly the folding of AMPs at membrane models. This research paves the way for future advances in quantum computing to enhance the accuracy and applicability of biomolecular simulations, with promising implications for the development of novel therapeutic agents.
\end{abstract}

\maketitle

\maketitle
\section{Introduction}
Antimicrobial peptides (AMPs) are critical components of the innate immune system present in all living organisms\cite{nayab2022review}. These peptides have been primarily associated with a defensive role against exogenous infections caused by bacteria, viruses, and fungi, and they are considered powerful and versatile endogenous antibiotics, capable of resisting bacterial adaptation for millions of years. However, recent research advances have pointed to the link between AMPs and a broader spectrum of diseases, such as cancer and various human inflammatory and autoimmune diseases, including aging\cite{jafari2022clinical, stuart2022regulatory}. Although more than 3,000 AMPs have been identified so far in distinct cells and tissues of animals, insects, plants, and bacteria, only a few have reached the pharmaceutical market\cite{dijksteel2021lessons}. Challenges for the clinical application of AMPs include cytotoxic effects, production  costs, and problems related to sustained, targeted, and effective delivery. The quest to discover and refine new antimicrobial peptides (AMPs), including both natural and engineered variants, represents a dynamic and promising field of research. The goal is to overcome these challenges, optimizing these peptides for medical use and leveraging their full potential as therapeutic agents.  Unraveling this matter is crucial when antibiotic resistance is a growing global threat but also in the fight against cancer and aging, areas where AMPs can still provide significant solutions. 

\begin{figure}[h]
\centering
\includegraphics[width=\columnwidth]{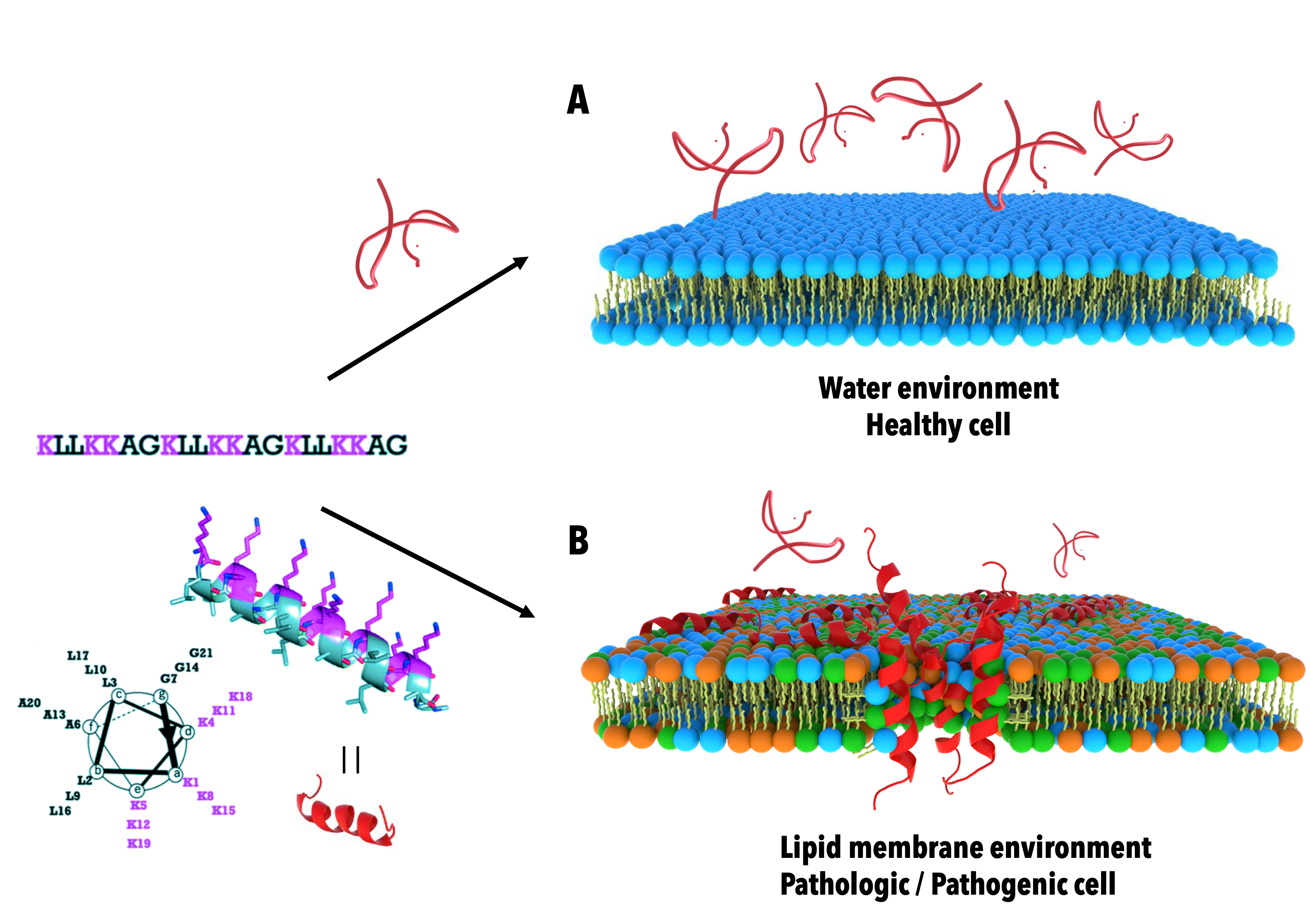}
\caption {\footnotesize Two scenarios of antimicrobial peptide (AMP) interactions with cell membranes. Panel A shows different AMP units in the presence of healthy mammalian cell membranes (the lipid head groups represented by blue spheres), where the peptides maintain a random coil conformation. Panel B depicts the interaction of AMPs with a pathogenic or pathological membrane model typical of bacterial membranes, some viruses, cancer cells, or senescent cells. Here, the AMPs adopt a helical conformation upon interacting with the membrane. The orange spheres represent the head groups of several anionic lipids commonly found in these altered membranes, highlighting the  structural adaptations of AMPs in different cellular environments.}
\label{fig:introduction}
\end{figure}

Despite varying length, sequence, and conformation, most AMPs share crucial structural and physicochemical properties: they are typically short, cationic, and amphipathic peptides. This unique combination of characteristics enables them to selectively target and interact with pathogenic or pathological membranes, such as those found in cancer, bacteria, and senescent cells. This selective targeting stems from a common feature in these membranes: a high proportion of negatively charged lipids, in contrast to what happens in healthy mammalian cells whose electrostatic charge density is normally negligible. AMPs are known to undergo conformational shifts, transitioning from random structures in solution to helical structures upon encountering a membrane, a change driven by their inherent amphiphilic nature (Fig. \ref{fig:introduction}). This transformation enhances the alignment of their hydrophobic dipole moments across the membrane, facilitating optimal interaction with the lipid bilayer. The spatial arrangement of the amino acid residues in AMPs is indispensable for their biological function. Following membrane binding, AMPs exert their effect through various mechanisms, including the barrel stave, carpet, and toroidal pore models, among other conformations\cite{wimley2010describing,li2017membrane}. A deeper understanding of these action mechanisms is essential to improve AMP design, moving from current trial-and-error methods towards more precise and effective strategies. For example, tuning the modeling is especially relevant when there is an alteration in the lipid composition. While there are models describing AMP interactions with cell membranes, comprehensive atomic-level details are scarce, indicating a need for more in-depth research in this area. 

The transformation of a polypeptide chain into its functional three-dimensional structure represents a central challenge in molecular biology, especially at interfaces such as the surface of a cell membrane or upon the influence of some heterogeneous environment. Despite their fundamental role in numerous biological mechanisms, the speed and dynamics of these folding processes remain puzzling.  The Levinthal paradox illustrates this complexity by highlighting the seeming impossibility of amino acid chains in finding their native, functional conformation in a biologically relevant
timescale if they were to explore all possible conformations \cite{zwanzig1992levinthal}. To address this challenge, a variety of computational and experimental approaches have been employed. AlphaFold initiative \cite{jumper2021highly} is a significant advance capable of predicting the three-dimensional structure of proteins with unprecedented accuracy. Nevertheless, this impressive technology is still unable of reliably predicting the interaction between the 3D structure of short peptides and the membrane models that account for specific lipid compositions. In parallel, molecular dynamics (MD) simulations have emerged as a powerful tool for investigating peptide and protein folding dynamics or at least structural stability under different conditions \cite{palmer2021molecular,suarez2022supepmem}. Standard MD simulations explore the energy landscape of the polypeptide chain, providing insights into the folding or structural evolution pathway. However, limitations in computational power often restrict the simulation timescales, hindering the observation of complete folding events or transitions between different states separated by significant energy barriers, even for relatively short sequences. Biased MD techniques overcome these limitations by nudging the simulation to sample diverse states\cite{simcock2021membrane,kabelka2021selecting}. These methods can significantly accelerate the folding process, allowing to study it with greater detail, including the presence of specific heterogeneous environments. Despite these advancements, predicting peptide and protein folding remains a complex task, particularly in the presence of membrane models due to their intricate interactions between the macromolecule and the lipid bilayer. This coupling underlines a critical need for enhanced methodologies that can accurately predict peptide structures as a function of their specific environment, which would allow significant advances in the characterization of known structures and further the development of new AMP candidates.

Thus, studying protein and peptide folding is an intrinsically very complex problem whose practical solution is beyond the reach of classical algorithms\cite{jumper2021highly,berger1998protein}. In this scenario, quantum computers emerge as a promising tool despite the noisy intermediate scale quantum (NISQ) era. Recent work has attempted to solve this problem for relatively short amino acid sequences within homogeneous media\cite{chandarana2023meta,robert2021resource,bopardikar2023approach}. These studies adopt several simplifying approaches that neglect specific chemical details, such as mapping amino acids onto single spheres and modeling their interaction energy using a simplified pairwise potential. Additionally, rotations of these spheres are limited to discrete angles relative to their nearest neighbors, further reducing computational complexity. Moreover,  underestimating explicit interactions with solvent molecules is another eventual source of imprecision. While these simplifications significantly improve computational efficiency and reduce the required number of qubits ($N_q$), they come at the cost of reduced accuracy. Nevertheless, these approaches offer a valuable tool for gaining initial insights into peptide structure, precisely in the quantum computing context, where computational resources are limited. While these methods remains refineable, there is room to make them more versatile. In particular, our focus on AMPs and their interaction with membrane models necessitates extending these approaches to incorporate a smooth interface between two media of differing polarities, trying to mimic the interface between a lipid bilayer and the aqueous phase in contact with it. In this scenario, the folding process becomes significantly more complex, as the inhomogeneous and anisotropic environment substantially influences the structure and function of AMPs. This attempt highlights the critical demand for intensified efforts in developing quantum computing techniques, potentially leading to breakthroughs in studying and designing novel AMPs.

Our work extends a quantum-computing routine for peptide folding in homogeneous media to predict the optimal structure of amino acid sequences at the transition region between hydrophilic and hydrophobic environments, used as membrane models. The original proposal of Robert et al. \cite{robert2021resource} demonstrated the effective use of quantum algorithms in optimizing the conformation of small peptides, employing a Hamiltonian (${\cal H}$) model for folding polymer chains on a lattice. This approach bridged the gap between simplified models and more detailed peptide representations.

Three amino acid sequences, chosen for their distinct characteristics: polar, non-polar, and having a high transversal hydrophobic dipolar moment when forming an alpha helix, were employed to test the new method in various homogeneous and non-homogeneous environments.
Our proposal introduces a valuable new dimension to existing computational models without adding substantial computational resource demands or unnecessary complexity. This represents a significant step towards refining more sophisticated and precise peptide modeling techniques, enhances our understanding of protein chemistry in complex environments and lays the foundation for future advancements in the field. We have confidence that this work will inspire further research, ultimately leading to the creation of robust peptide structures that effectively consider different environmental conditions. This expansion of scientific knowledge holds promise for therapeutic applications, harnessing the unique capabilities of quantum computing to explore the intricate details of protein structures.

\section{Materials and methods}
\subsection{Interface implementation}
\subsubsection{Background}
\begin{figure}[h]
\centering
\includegraphics[width=\columnwidth]{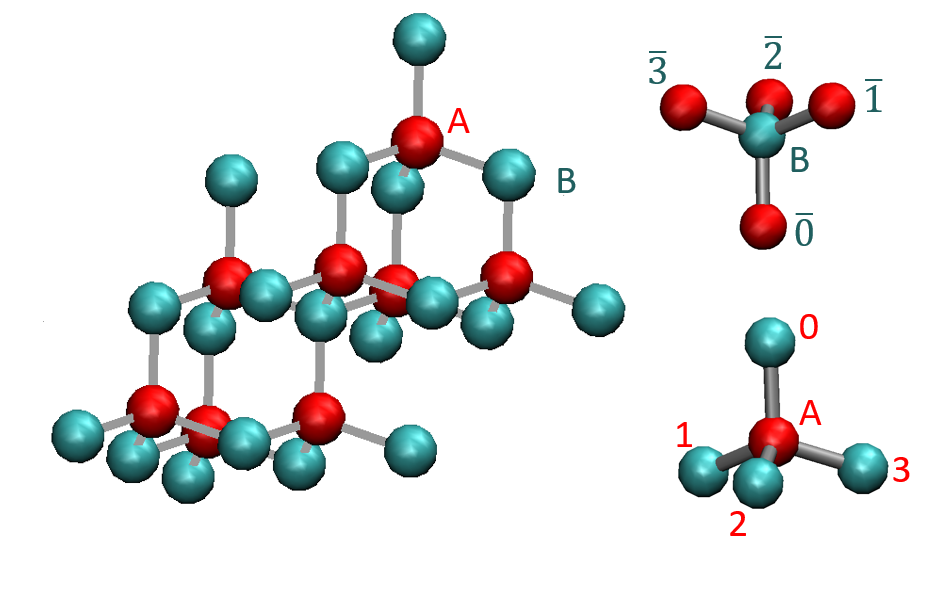}
\caption {\footnotesize Tetrahedral lattice in which a representative peptide will move, with sublattices A (red) and B (green) indicated.}
\label{fig:diamond}
\end{figure}

The prediction of peptide structure in homogeneous media within the \texttt{protein\_folding} module of the \texttt{qiskit\_research}\cite{the_qiskit_research_developers_and_contr_2023_7776174} library utilizes a quantum computational approach\cite{robert2021resource} that employs a model Hamiltonian and a variational quantum algorithm to fold a polymer chain on a tetrahedral lattice. The Hamiltonian is based on the pairwise Miyazawa-Jernigan (MJ) potential\cite{Miyazawa1985,Miyazawa1996}, where each amino acid is represented by a single sphere. The MJ coarse-grained representation ignores chemical details but it is expected to describe reasonably well the intramolecular interactions between the amino acid residues. The lattice model simplifies the representation of the peptide to make it computationally feasible for quantum simulations. Specifically, two sets of non-equivalent lattice points (\textbf{A} and \textbf{B}) are defined as sublattices. At sites \textbf{A}, the polymer can only grow in the directions $t_{i} \in \{0,1,2,3\}$ while at site \textbf{B}, the possible directions are $t_{i} \in \{\bar{0},\bar{1},\bar{2},\bar{3}\}$ (Fig. \ref{fig:diamond}). Throughout the sequence, the \textbf{A} and \textbf{B} sites alternate, allowing us to adopt the convention that \textbf{A} and \textbf{B} sites correspond to even and odd  values of \textit{i}, respectively. Without loss of generality, the first two turns can be set to $t_{1}=\bar{1}$ and $t_{2}=0$ due to symmetric degeneracy. The turns are encoded by assigning a combination of two qubits per axis. Each pair of qubits can be in one of four possible states: 00, 01, 10 and 11, thus allowing for a precise and efficient encoding of turns.

Therefore, a bitstring represents the three-dimensional structure of the peptide, which codifies the sequential turns of the coarse-grain beads.  A relatively low number of conformation and interaction qubits is required under this approach, including penalty terms to prevent meaningless conformations of the peptide, such as residue overlaps and chiral violations. The number of qubits required for this model scales quadratically with the number of amino acid residues in the peptide sequence $(N)$ while the number of terms in the Hamiltonian scales in $O(N^4)$. Adding sidechains and incorporating state-of-the-art classical force fields based on Lennard-Jones and Coulomb interactions is also possible by keeping the structure of the employed Hamiltonian, albeit this would require a higher number of particles and so a higher number of qubits.

\begin{figure}[h]
\centering
\includegraphics[width=\columnwidth]{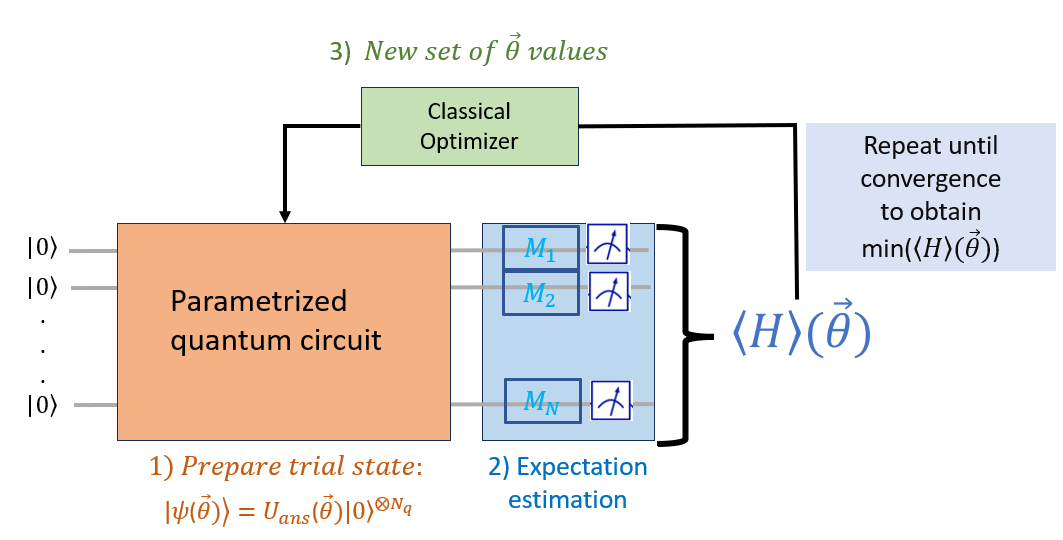}
\caption {\footnotesize Schematic representation of how the VQE algorithm works.}
\label{fig:vqe}
\end{figure}

Since the \texttt{protein\_folding} module takes advantage of a Variational Quantum Eigensolver (VQE)\cite{tilly2022variational} (Fig. \ref{fig:vqe}), the Hamiltonian is minimized for each iteration of the  parametrized quantum circuit. This means that the Hamiltonian should be self-consistent to be executed in the quantum processor unit without depending on the state of the qubits.

\subsubsection{Amino acids location concerning the interface}
\begin{figure}[h]
\centering
\includegraphics[width=\columnwidth]{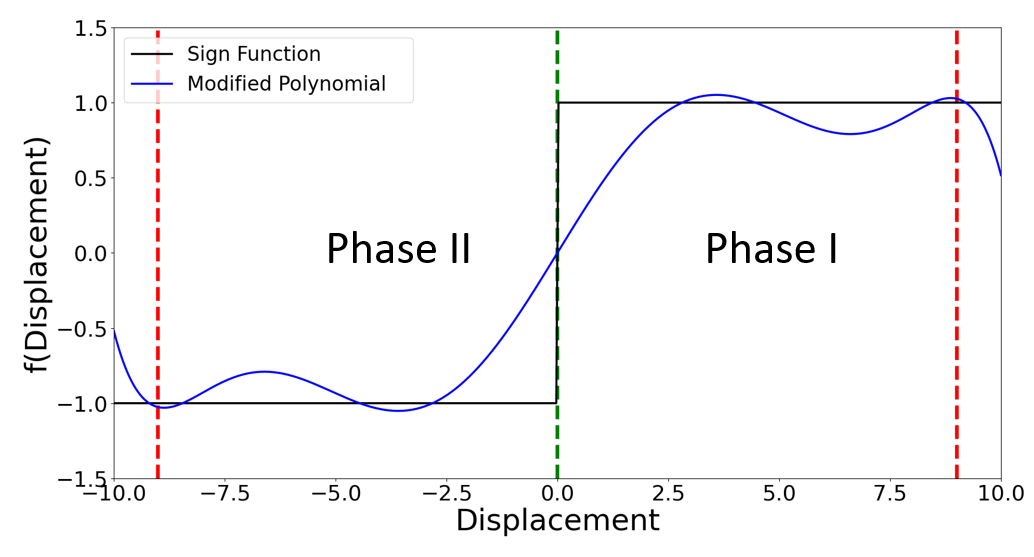}
\caption {\footnotesize Comparison between the sign function (black line) and its 7th-degree polynomial approximation (blue line). The vertical green line indicates the location of a virtual plane separating both solvents and the dashed red lines indicate the distances beyond which the difference between the sign function and its polynomial approximation diverges.}
\label{fig:signfunction}
\end{figure}

The positional displacement of each bead along a specific tetrahedral axis $a$ is quantified as:

\begin{equation}\label{eqdisplacement}
    \Delta n_{a}(j)=\sum_{k=1}^{j-1}(-1)^{k}f_{a}(k) + \Delta_{a}
\end{equation}

\noindent where the sum is performed from the first to the current bead ($j$), $\Delta_{a}$ represents the distance along axis $a$ from the first bead of the peptide (which is always fixed in our approach) to the phase-separating plane. The function $f_a(k)$ returns $1$ if there is a displacement along axis $a$ for the turn of amino acid $k$, and $0$ otherwise. The term $(-1)^k$ indicates the directionality of the turn relative to sublattices \textbf{A} or \textbf{B}, effectively showing whether the movement brings the bead closer to or further from the phase-separating plane. $\Delta n_{a}(j)$ inherently determines the phase location of amino acid $j$, as well as the distance to the phase-separating plane. It is important to note that the contribution of the interaction between the amino acids and its corresponding phase cannot be proportional to $\Delta n_{a}(j)$ since, in that case, such interaction would linearly increase the affinity or repulsion (depending on the sign of the interaction) of each amino acid to each phase as a function of the distance to it. On the other hand, extracting directly the sign of this function is not a trivial task without reading the state of the qubits. While auxiliary qubits could facilitate this, they would also increase the computational demands, which is inconvenient. Additionally, directly using a step function to identify the location of the bead at each medium would be an unsuitable approach since actual interfaces, such as that between an aqueous media and a lipid membrane, are smooth. The roughness of such interfaces is comparable to the diameter of a water molecule (3-6 \r{A}), as estimated from neutron reflectometry analysis\cite{PhysRevLett.67.2678}, so a gradual transition between both phases is foreseeable. To address all these issues, we decided to use a polynomial approximation to the sign function (see Fig. \ref{fig:signfunction}) as a scaling factor for the Hamiltonian contribution of the interaction between each amino acid and the corresponding medium: 

 \begin{equation}\label{eq:polynolmial}
 \begin{split}
 &f(x) = 0.48175x - 0.0182x^{3} + \\ 
 & +(2.95 \cdot 10^{-4}) x^{5}- (1.56 \cdot 10^{-6}) x^{7}
 \end{split}
 \end{equation}

This approach provides a smooth transition between the two phases, with an interface thickness of approximately 5 arbitrary units, corresponding to the distance between two beads in the peptide's lattice. From this region and at distances lower than 9 units, the polynomial function exhibits relatively small oscillations. At longer distances this polynomial diverges from the sign function. Since the studied peptides are quite short (maximum 10 amino acids) and unlikely to extend further from the interface center, this approximation considerably tailors our purposes.

\subsubsection{Tuning the interaction between amino acids as a function of the media}
The approach already implemented in the \texttt{qiskit\_research} library is well-designed for modeling peptide folding in homogeneous media. Details of the implementation are well documented in previous publications\cite{robert2021resource} as well as in the official repository of the \texttt{protein\_folding} module \cite{the_qiskit_research_developers_and_contr_2023_7776174}. However, its applicability to functional antimicrobial peptides is limited, as these peptides exert their biological function by interacting with the surface of pathological membranes, which could be roughly modeled as a hydrophilic/hydrophobic interface. Thus, several modifications were introduced into the original model. First, the Miyazawa-Jernigan (MJ) parameters were modified following the work of Leonhard et al.\cite{Leonhard2004,leonhard2006three} to account for interactions between residue beads in different phases:

\begin{equation}
e_{i,j}^{Leonhard}=e_{i,j}^{MJ}-e_{i,phase}-e_{j,phase}
\end{equation}\\
where $e_{i,j}^{MJ}$ is the original value of the MJ interaction terms between amino acids $i$ and $j$ and  $e_{k,phase}$ (with $k$ = $i$ or $j$) represents the interaction of amino acid $k$ with the phase it resides in. The value of $e_{k,phase}$ for a homogeneous phase is calculated using the following equation:

\begin{equation}
\begin{split}
e_{k,phase}&=\frac{1}{2}(1-C_{s})e_{kk}^{MJ}+\bar{\omega}+\frac{C_{s}}{2n}\sum_{i=1}^{20}e_{ii}^{MJ}\\&= \frac{1}{2}(1-C_{s})e_{kk}^{MJ}+\bar{\omega}'
\end{split}
\end{equation}\label{Leonhard}

\noindent where $i$ iterates over the total number of amino acid types, and $C_{s}$ determines the contrast between phases. A positive $C_{s}$ favors contact between solvent and hydrophilic residues, while a negative $C_{s}$ favors contact between solvent and hydrophobic residues. $\bar{\omega}$ determines the average interaction between amino acids and the solvent. Negative $\bar{\omega}$ indicates attraction, while positive $\bar{\omega}$ indicates repulsion. At the transition region between two phases of different polarity $e_{k,phase}$ will be replaced by $e_{k,phase'}$:

\begin{equation}
\begin{split}
e_{k,phase'}=\frac{1}{2} &\left[(1-\hat{S}) \cdot e_{k,phase_1} +  \right.\\
&\left. +(1+\hat{S}) \cdot e_{k,phase_2}\right]
\end{split}
\end{equation}

\noindent where $\hat{S}$ can take values between $1$ and $-1$, depending on whether the amino acid is in the polar or nonpolar phase. In our case, $\hat{S}$ will be replaced by the function provided by equation \ref{eq:polynolmial}. Depending on the value of $\hat{S}$, $e_{k,phase'}$ can be closer to the value of $e_{k,phase1}$ or $e_{k,phase2}$. This term can be switched off in the Hamiltonian, in case the study is performed in an homogemeous media and so the original MJ potential is employed, as it is a boolean parameter.

\subsubsection{Interfacial contribution to the Hamiltonian}
The previous modifications of the MJ potential account for the different ocurring interactions between amino acids based on their location within the aqueous or membrane phases. Besides interacting with each other, amino acids also directly interact with the solvent in both media. Thus, a new contribution, ${\cal H}_{sol}(q_{cf})$, has been added to the total Hamiltonian:

\begin{equation} \label{eq:hamiltonian}
{\cal H}(q)={\cal H}_{gc}(q_{cf})+{\cal H}_{ch}(q_{cf})+{\cal H}_{in}(q)+{\cal H}_{sol}(q_{cf})
\end{equation}

\noindent where $q={q_{cf}, q_{in}}$ represents the complete set of qubits used in the model, including both the conformation qubits ($q_{cf}$) and the interaction qubits ($q_{in}$). The first three-terms description is available in \cite{robert2021resource,chandarana2023meta}. Briefly:

\begin{itemize}
    \item ${\cal H}_{gc}(q_{cf})$ accounts for the geometrical constraints imposed by the tetrahedral lattice structure of the amino acids.
    \item ${\cal H}_{ch}(q_{cf})$ enforces the correct stereochemistry of the sidechains (when present), ensuring the accuracy of the amino-acid-chirality representation.
    \item ${\cal H}_{in}(q)$ accounts for the interactions between neighboring beads using the Miyazawa-Jernigan (MJ) potential. 
\end{itemize}

The new term ${\cal H}_{sol}(q_{cf})$ accounts for the interaction between the amino acids and each solvent. This term has been defined here as:

\begin{equation} \label{eq:Hsol}
{\cal H}_{sol}(q_{cf}) = \sum_i{\Delta P \cdot \gamma_i \cdot \hat{S}}
\end{equation}

\noindent $\Delta P$ represents the polarity difference between the two media and $\gamma_i$ represents a quantitative measurement of the hydrophobicity, or affinity of each residue for a hydrophobic media.  In the present work, the parameters used were proposed by Fauchere and Pliska \cite{fauchere1983hydrophobic}  (see Table  \ref{Table:1}), although there are different proposals for this parameter in the literature, obtained from a variety of methods \cite{kyte1982simple,wimley1996experimentally,hessa2005recognition,moon2011side}.

Note that charged residues, with the strongest attraction for polar solvents, have the lowest (most negative) $\gamma$ values; polar residues exhibit moderate values depending on their specific side chains, ranging from slightly negative to slightly positive; nonpolar residues have consistently positive values; while aromatic residues, with their large hydrophobic rings, possess the highest positive values of $\gamma$. The final expression for ${\cal H}_{sol}(q_{cf})$ provides a negative contribution to the Hamiltonian, favors the interaction of amino acids with $\gamma < 1$ (mainly nonpolar and aromatic) at the positive side of the interface (phase I) and for amino acids with $\gamma > 1$ (mainly charged) at the negative side of the interface (phase II) if $\Delta P >1$. The higher the value of $\Delta P$ the stronger this contribution. 

Both the pairwise MJ potential and the hydrophobicity $\gamma$  have arbitrary units and both are of the same order, so they compete with each other to modulate the optimal structure of the peptide at the interface. Importantly, this Hamiltonian implementation does not require additional qubits, and the number of extra operations is modest. In particular, the calculation for a sequence of 10 amino acids the number of required qbits is $N_q=22$. Hence, including the extra dimension of the interface, the final computational cost is not remarkably higher compared to the original model for homogeneous media.
\begin{table}[h]
\caption{Fauchere and Pliska \cite{fauchere1983hydrophobic} hydrophobicity scale.}\label{Table:1}%
\begin{ruledtabular}
\begin{tabular}{@{}ccc@{}}
\textbf{Amino acid} & \textbf{$\gamma$} & \textbf{Residue type} \\
\hline

ASP&  $-0.77$ &  Charged ($-$)\\
GLU &   $-0.64$ & Charged ($-$)\\
LYS &  $-0.99$ & Charged ($+$)\\ 
ARG &  $-1.01$ & Charged ($+$) \\ 
HIS &   $0.13$ & Charged ($+$) \\ 
GLY &  $0.00$ & Nonpolar\\ 
ALA &  $0.31$ & Nonpolar\\ 
VAL &  $1.22$ & Nonpolar \\ 
LEU &  $1.70$ & Nonpolar\\ 
ILE &  $1.80$ & Nonpolar\\ 
PRO &  $0.72$ & Nonpolar\\ 
MET &  $1.23$ &  Nonpolar\\ 
PHE &  $1.79$ & Aromatic\\ 
TRP &  $2.25$ & Aromatic\\ 
TYR &  $0.96$ &  Aromatic\\ 
THR &  $-0.04$ & Polar \\ 
SER &  $0.26$ & Polar \\ 
CYS &  $1.54$ & Polar \\ 
ASN &  $-0.60$ & Polar\\ 
GLN &  $-0.22$ & Polar\\ 
\end{tabular}
\end{ruledtabular}
\end{table}

All the described modifications to the model were implemented in the \texttt{protein\_folding} module of the \texttt{qiskit\_research} library. The whole code is written in Python \cite{van1995python}, making special use of the Qiskit \cite{Qiskit}, Numpy \cite{Harris2020}, Matplotlib \cite{Hunter2007} and Mayavi \cite{ramachandran2011mayavi} libraries and it is publicly available at 
\href{https://github.com/TeamMduse/Quantum-Computing-in-Peptide-Folding-Targeting-Cellular-Membranes}{https://github.com/TeamMduse}.

\subsection{Studied systems and parameters}
We have employed three amino acid sequences denoted \textbf{P1}, \textbf{P2}, and \textbf{P3} to validate our approach. These sequences were chosen to exhibit distinct affinities for media of differenbt polarity, based on the $\gamma$ values presented in Table \ref{Table:1}.
\begin{enumerate}
    \item \textbf{P1} comprises exclusively hydrophobic amino acids (Leucine and Tryptophan), maximizing its affinity for nonpolar environments.
    \item \textbf{P2} consists solely of charged amino acids (Glutamic Acid and Arginine), promoting its interaction with polar media.
    \item \textbf{P3} represents a more intricate sequence, containing charged amino acids of opposing charges (Glutamic Acid and Arginine), highly polar and neutral residues (Serine), highly polar (Tryptophan) and neutral-nonpolar (Glycine) amino acids, distributed such that generates a significant transversal component of the hydrophobic dipole moment when adopting a helical conformation.
\end{enumerate}

    \begin{figure}[h]
        \centering
        \includegraphics[width=8cm]{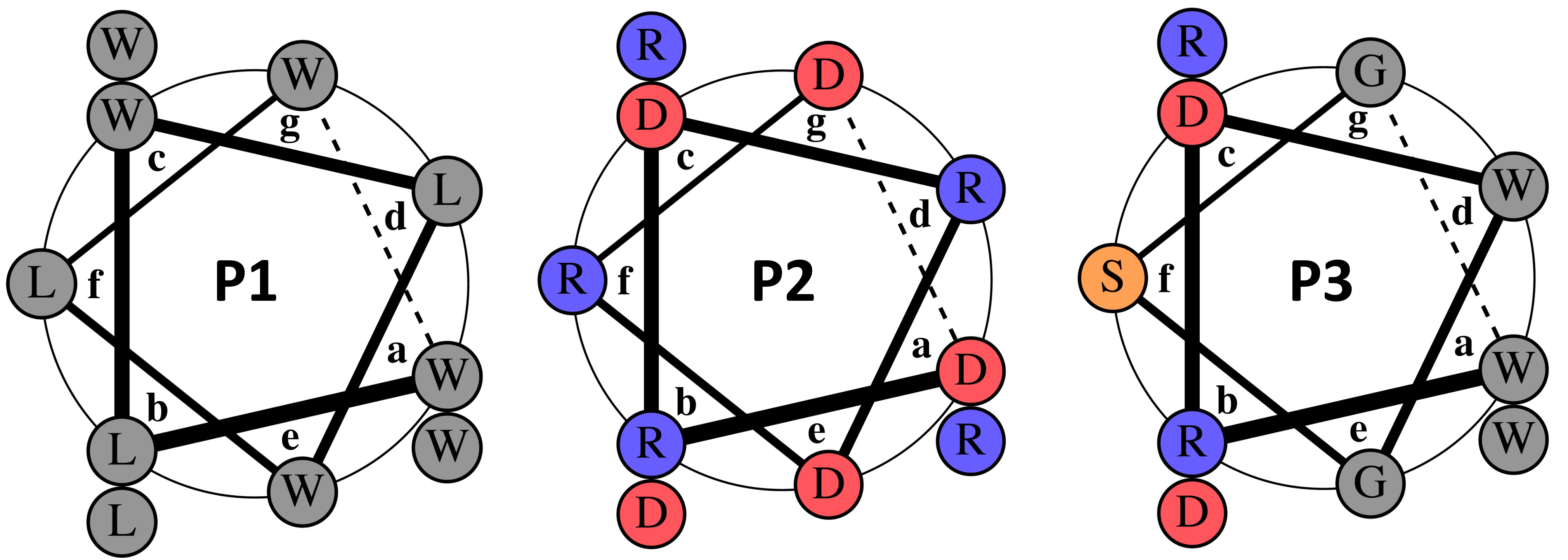}
        \caption {\footnotesize Wheel representation of \textbf{P1}, \textbf{P2}, and \textbf{P3} with sequences \textbf{WLWLWLWWLW}, \textbf{DRDRDRDRDR} and \textbf{WRDWGSGWDR}, respectively. W, L, R, D, G and S denote Tryptophan, Leucine, Argininie, Glutamic Acid, Glycine and Serine, respectively. \textbf{P1} and \textbf{P2} are expected to exhibit a high affinity for a nonpolar and for a polar media, respectively, while \textbf{P3} is expected to acquire a helical conformation at the interface between two media of different polarity, with a high transvesal component of the hydrophobic dipolar moment. Positively charged amino acids (Arginine) are in blue, negatively charged amino acids (Glutamic Acid) are in red, polar amino acids (Serine) are in orange, and neutral-nonpolar (Glycine and Leucine) and aromatic (Tryptophan) amino acids are in grey.}
        \label{fig:wheels}
    \end{figure}

This selection of sequences allows for a comprehensive evaluation of the ability of our approach to accurately capture and predict the behavior of peptides with a range of physicochemical properties in environments of differing polarity. By analyzing their partitioning between phases of opposite polarity under the influence of our model, we can assess its efficacy in reflecting the underlying principles of amino acid–solvent interactions.

It is worth reminding that the model output is a bitstring with the sequential turns of the amino acids relative to their previous closest neighbors. Substantially, the first two beads, representing the first two amino acids, have fixed positions. The location of these beads defines their distance and orientation concerning the plane separating the two phases. Upon these restraints and those provided by the model (chemical consistency and tetrahedral lattice), the turns of the remaining beads establish the structure of the peptide. Axis 1 of the tetrahedral lattice (see Fig. \ref{fig:diamond}) was chosen to define the polarity gradient. The first bead of the peptides, representing the first amino acid, was placed at different positions along the same axis ($-1$, $-0.5$, 0, 0.5, and 1).  This set of configurations led to various  distances between such a bead and the phase-separating plane towards both solvents. In all cases, the second bead was aligned along the same axis in the direction of the more polar solvent. Additionally, different values of $\Delta P$ (0.1, 1 and 10) were essayed in order to balance the competition between the weight of the interaction between amino acids and the weight of their interaction with the solvent. Finally, the weights of the existing penalty terms in the original approach were increased from 10 to 1000. This was done to prevent them from being overshadowed by the new contribution to the Hamiltonian. The results obtained from this combination of parameters for the three studied sequences are presented in the next section.

\section{Results and Discussion}
The optimal conformations of the peptide sequences described in the methods section were obtained throughout the minimization of the Hamiltonian, using the VQE algorithm, under different conditions: in polar and nonpolar homogeneous phases as well as at polar/nonpolar interfaces. The location and orientation of the first two amino acids of each sequence concerning the phase-separating plane (when two different media are considered) were restrained. 

\subsection{Homogeneous media}
\begin{figure}[h]
    \centering
    \includegraphics[width=\columnwidth]{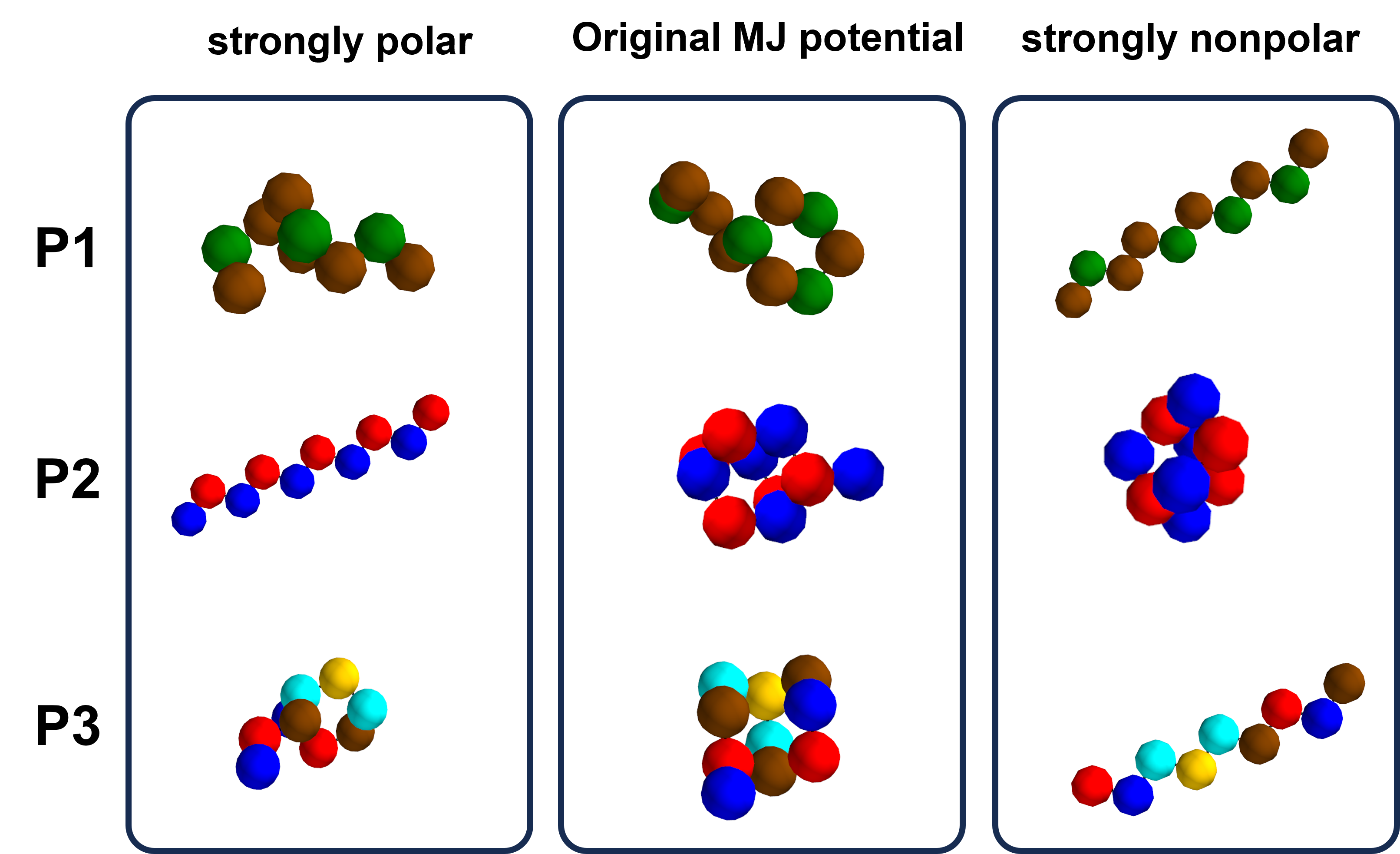}
    \caption {\footnotesize Conformations of peptides \textbf{P1}, \textbf{P2}, and \textbf{P3} in homogeneous media of different polarity and also ignoring the polarity of the media (i.e. using the original MJ potential). Different colors are employed for each residue: W in brown, L in green, R in blue, D in red, G in cyan, and S in yellow.}
    \label{fig:homogeneous}
\end{figure}

The conformation of each peptide is highly sensitive to the polarity of the environment in homogeneous media (Fig. \ref{fig:homogeneous}). The peptide consisting just of hydrophobic amino acids (\textbf{P1}) and that formed just by charged amino acids (\textbf{P2}) exhibit an opposite behavior, as expected. \textbf{P1} is folded in polar environments and fully extended in nonpolar media, while \textbf{P2} is fully extended in polar environments and folded in nonpolar media. We observe that the behavior of \textbf{P3} is similar to that of \textbf{P1}. The three peptides acquire different folded conformations when using the original MJ potential.

\noindent The extended conformations of \textbf{P1} and \textbf{P2} in nonpolar and polar media, respectively, are due to highly favorable interactions between the amino acids consisting of those peptides and the model solvent in those scenarios. In these cases the interaction with the media largely dominates the Hamiltonian while the intramolecular interactions are less important. Conversely, the folded conformations of the same peptides in the opposite media (\textbf{P1} in polar solvent and \textbf{P2} in nonpolar solvent) arise from the favorable interactions between the amino acids consisting of those peptides combined with unfavorable interactions with the media. No clear secondary structure patterns are observed in any of the folded conformations.

\subsection{Polar/nonpolar interfaces}
According to the presented calculations within two media, peptide conformation relies heavily on several competing energy factors.  These factors primarily include the interactions among the amino acids, as well as their interactions with the two model solvents used in the study.
For instance, it is possible to identify scenarios where two different amino acids have a strong mutual attraction but an even stronger affinity for opposite phases. This disbalance can definitely influence the peptide conformation, potentially resulting  in the spatial separation of these amino acids despite their intrinsic attraction. The emergence of disparate peptide configurations in heterogeneous environments highlights the complex interplay between intra-peptide and peptide-solvent interactions and the relevance of implementing an interface
model.

\begin{figure}[h]
    \centering
    \includegraphics[width=\columnwidth]{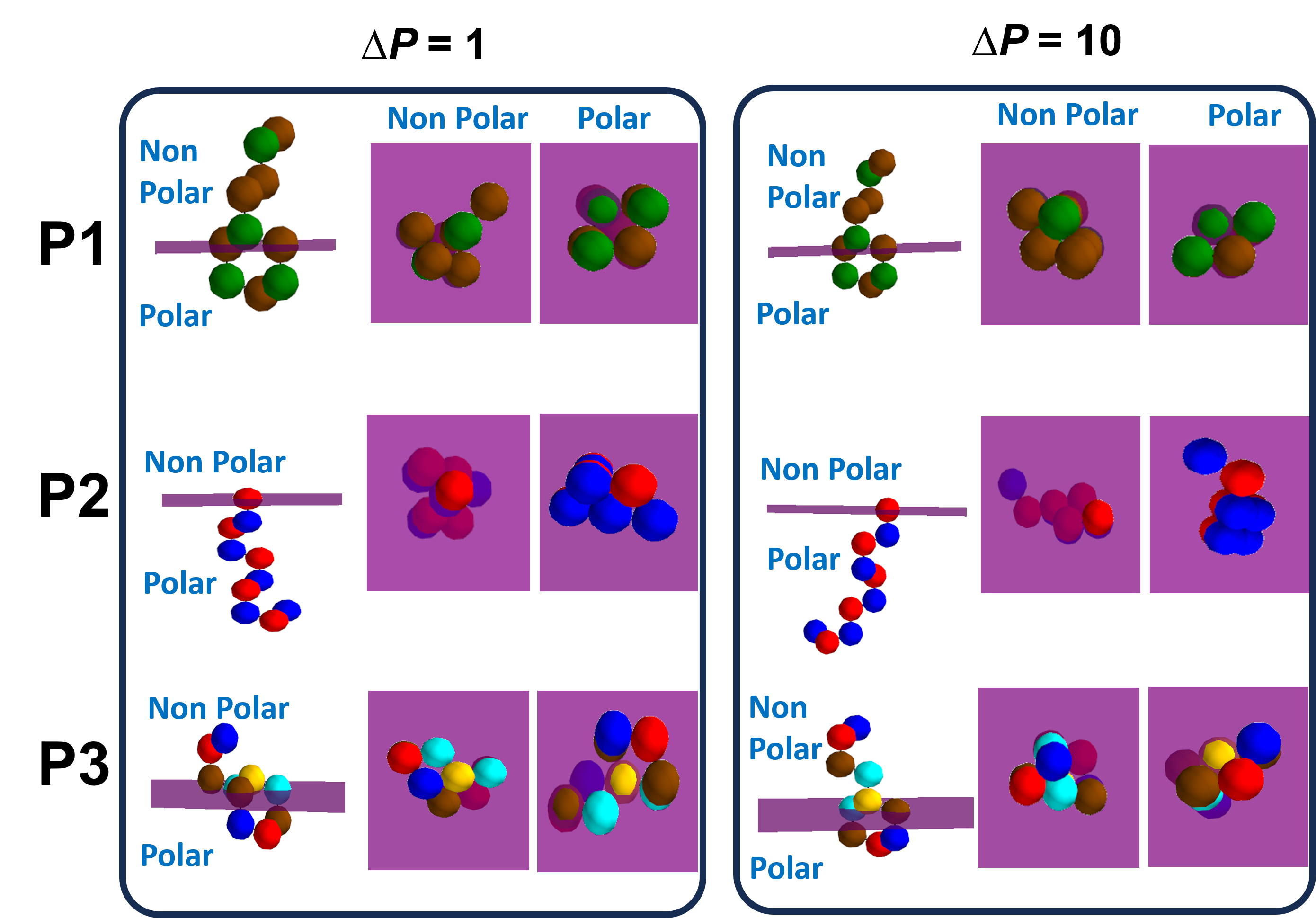}
    \caption {\footnotesize Conformations of peptides \textbf{P1}, \textbf{P2}, and \textbf{P3} at the interface between two media of different polarity. For each peptide a side view (left), a view from the nonpolar phase (middle) and another view from the polar phase (right), are shown. In the side view the interface is shown as an horizontal line. Different colors are employed for each residue: W in brown, L in green, R in blue, D in red, G in cyan, and S in yellow.}
    \label{fig:heterogeneous}
\end{figure}

The results observed for sequences \textbf{P1}, \textbf{P2}, and \textbf{P3} clearly show that hydrophobic residues are more stable in the nonpolar environment, even if the peptide sequence needs a turn to reorientate the corresponding coarse-grained beads. Note that the location of the first two amino acids is fixed in our approach, so the peptide cannot travel as a whole from one media to the other, and the orientation of the first two amino acids concerning the phase-separating plane is not optimized by minimizing the Hamiltonian. The behavior of charged residues (D and R) is opposite to that of hydrophobic residues (W and L). Thus, the conformation of peptides \textbf{P1} and \textbf{P2} at the interfacial model could be easily predicted (Fig. \ref{fig:heterogeneous}). Besides, \textbf{P3} was designed to ideally fold into a helical structure in this heterogeneous environment. This peptide does not exhibit a clear trend to stay in one or other phase, but the amino acids are distributed between the two media, as expected. The obtained conformation is not an ideal helix. Moreover, some amino acids are located in the wrong phase, probably due to the limitations of the employed tetrahedral model. While the possibility of convergence to local minima in the VQE algorithm cannot be completely ruled out, we took thorough measures to mitigate this issue. The calculations were repeated multiple times for the most controversial cases, employing a conservatively high number of iterations and varying the seeds, yet these adjustments did not alter the final structure obtained.

 Additional calculations using different initial coordinates for the two first amino acids of the employed sequences provided different conformations of the peptides (as expected) but they followed the same qualitative behavior as the results shown in Fig. \ref{fig:heterogeneous}.

\section{Conclusions}
This study aims to contribute significantly to the field of peptide folding simulations using quantum computing by introducing a new dimension to a preexisting model \cite{robert2021resource}. Our research focuses on the folding of peptides in different environments, particularly at the interface between hydrophobic and hydrophilic phases, which is critical for understanding the function of antimicrobial peptides (AMPs) in biological systems. Based on a modified version of the Miyazawa-Jernigan potential, our approach employs a tetrahedral lattice model to represent peptide structures, combined with the introduction of a Hamiltonian  contribution accounting for the interaction between the amino acids and the solvent in each phase. The transition region from one to another media is modeled as an smooth function, trying to mimic the actual interface at the vicinity of a cell membrane. Furthermore, our implementation is computationally efficient and does not require additional qubits compared to the original model that only considers an homogeneous phase. Our findings demonstrate that peptides exhibit distinct folding patterns in response to the polarity of their surrounding environment. Results point out the potential of quantum computing to simulate complex biological processes, which classical computing approaches struggle to accomplish due to computational limitations. 

While integrating a polar/nonpolar interface in peptide folding represents a significant achievement, the extended model leans on approximations originally proposed for calculations in homogeneous media. In particular, the consideration of a tetrahedral lattice that restrains the turns of the amino acids combined with the minimalist MJ pairwise potential interaction seems to be inaccurate in successfully predicting peptide secondary structure. The limited number of available qubits currently makes it unfeasible to add more degrees of freedom and a more reliable potential for amino acid interaction.

The specific aim of this study is to introduce, for the first time, an efficient method to leverage quantum computing for predicting reasonable peptide structures at the interface between media of different polarity. This starting point opens new avenues for understanding peptide interactions at the molecular level, which could lead to significant advances in developing new therapeutic agents, particularly in the realm of antimicrobial peptides. Future research should aim to refine the quantum computational approach to enhance its accuracy and applicability to a broader range of biomolecules. Furthermore, integrating more detailed chemical properties and interactions into the model could yield even more nuanced insights into peptide folding dynamics. The general goal is to develop a quantum computational framework capable of simulating various biological processes. Advancing our understanding and capabilities in molecular biology underscore the critical importance of ongoing research and development in the field of quantum computing, particularly in its application to complex biological systems.


\begin{acknowledgments}
D.C.T thanks to the Ministerio de Universidades for his predoctoral contract (FPU22/00636). This work was supported by the European Union’s Horizon Europe research and innovation programme under the Marie Sklodowska-Curie grant agreement Bicyclos N° 101130235, by the Interreg Sudoe and the ERDF (S1/1.1/P0033), by the Spanish Agencia Estatal de Investigación (AEI) and the ERDF (PID2022-141534OB-I00, PDC2022-133402-I00, CNS2023-144353 and PID2019-111327GB-I00), by Xunta de Galicia and the ERDF (ED431C 2021/21, ED431B 2022/36, 06\_IN606D\_2021\_2600276, 02\_IN606D-2022-2667887 and Centro singular de investigación de Galicia accreditation 2016-2019, ED431G/09 and Axencia Galega de Innovación through the Grant Agreement “Despregamento dunha infraestrutura baseada en tecnoloxías cuánticas da información que permita impulsar a I+D+I en Galicia” within the program FEDER Galicia 2014-2020. Simulations on this work were performed using the Finisterrae III Supercomputer, funded by the project CESGA-01 FINISTERRAE III.
\end{acknowledgments}

\bibliography{lib}

\end{document}